\documentclass{PoS}
\usepackage{url}
\usepackage{graphicx}
\usepackage{wrapfig}
\usepackage{eufrak} 
\usepackage{amsmath}   
\usepackage{microtype}
\usepackage{times}
\usepackage[T1]{fontenc}
\usepackage{amssymb}
\usepackage{mathptmx}
\usepackage{color}
\let\normalcolor\relax 
\newcommand{\bea}{\begin{eqnarray}}
\newcommand{\eea}{\end{eqnarray}}
\newcommand{\bqa}{\begin{eqnarray}}
\newcommand{\eqa}{\end{eqnarray}}

\newcommand{\eps}{\varepsilon}
\newcommand{\beq}{ \begin {equation} }
\newcommand{\eeq}{\end{equation}}
\newcommand{\be}{\begin{eqnarray}}
\newcommand{\ea}{\end{eqnarray}}
\newcommand{\nn}{\nonumber}

\title{
{\small
\texttt{DESY 10-073
\\
HEPTOOLS 10-020
\\
SFB/CPP-10-43\\[1.5cm]
}}
Some variations of the reduction of
one-loop Feynman tensor integrals}

\ShortTitle{Reduction of tensor integrals}

\author{Jochem Fleischer
\\
Fakult\"at f\"ur Physik, Universit\"at Bielefeld, Universit\"atsstr. 25,  33615
Bielefeld, Germany
\\
        E-mail: \email{Fleischer@physik.uni-bielefeld.de}}

\author{\speaker{Tord Riemann}
        \\
        Deutsches Elektronen-Synchrotron, DESY, Platanenallee 6, 15738 Zeuthen, Germany
        \\
        E-mail: \email{Tord.Riemann@desy.de}}

\abstract{%
We present a new algorithm for the reduction of one-loop \emph{tensor}  Feynman integrals with $n\leq 4$ external legs to  \emph{scalar} Feynman integrals $I_n^D$ with $n=3,4$ legs in $D$ dimensions, where $D=d+2l$ with integer $l \geq 0$ and generic dimension $d=4-2\epsilon$, thus avoiding the
appearance of inverse Gram determinants $()_4$.
As long as $()_4\neq 0$, the integrals  $I_{3,4}^D$ with $D>d$ may be further expressed by the usual dimensionally regularized  scalar functions $I_{2,3,4}^d$.
The integrals  $I_{4}^D$ are known  at $()_4 \equiv 0$, so that we may
extend  the numerics to small, non-vanishing $()_4$ by applying a dimensional recurrence relation.
A numerical example is worked out.
Together with a recursive reduction of 6- and 5-point functions, derived earlier, the calculational scheme allows a  stabilized reduction of $n$-point functions with $n\leq 6$ at arbitrary phase space  points.
The algorithm is worked out explicitely for tensors of rank $R\leq n$.
}

\FullConference{13th International Workshop on Advanced Computing and Analysis Techniques in Physics Research, ACAT2010\\
		February 22-27, 2010\\
		Jaipur, India}

\begin{document}

\section{Introduction\label{s-intro}}
The efficient and stable evaluation of tensor Feynman integrals,
\bea
\label{definition}
 I_n^{\mu_1\cdots\mu_R} &=&  ~ C(\varepsilon) ~\int \frac{d^d k}{i\pi^{d/2}}~~\frac{\prod_{r=1}^{R} k^{\mu_r}}{\prod_{j=1}^{n}c_j^{\nu_j}},
\eea
with denominators $c_j$, having \emph{indices} $\nu_j$ and \emph{chords}
$q_j$,
\begin{eqnarray}\label{propagators}
c_j &=& (k-q_j)^2-m_j^2 +i \epsilon ,
\end{eqnarray}
is an important ingredient of precision calculations for collider physics.
The normalization $C(\varepsilon)$ plays a role for divergent integrals only and is conventional,
$C(\varepsilon) = (\mu)^{2\eps} \Gamma(1 - 2\eps) / [\Gamma(1 + \eps) \Gamma^2(1 - \eps)]$.
Here, we use the generic dimension $d=4-2\epsilon$ and $\mu=1$.

For $n\leq 4$, the problem was basically solved in the Seventies of the last century \cite{'tHooft:1979xw,Passarino:1978jh} by tensor reduction with an ansatz of linear equations in terms of scalar integrals, and their evaluation in terms of logarithms and dilogarithms.
 For predictions of massive particle production at the LHC or ILC, one needs multi-dimensional phase space integrals over typically  hundreds to thousands of Feynman diagrams  with $n\leq 6$ external legs and tensor ranks $R\leq n$.
Further, the so-called $\epsilon$-expansion is needed in higher-order calculations.
Over the years, a variety of papers appeared, and a comprehensive survey is well beyond the scope of this contribution.
One of the approaches is purely algebraic and rests on the representation of single tensor integrals by scalar integrals, where the latter are defined in higher dimensions $D=d+2l$ and may have also higher indices $\nu_j+m$ \cite{Davydychev:1991va}, e.g.:
\begin{eqnarray}
\label{tensor1}
  I_n^{\mu} & =&  \int \frac{d^d k}{{i\pi}^{d/2}} k^{\mu} \prod_{r=1}^{n} \, {c_r^{-1}}  = -\sum_{i=1}^{n} \, q_i^{\mu} \, I_{n,i}^{d+2} ,
\\
\label{tensor2}
 I_{n}^{\mu\, \nu}& =& \int \frac{d^d k}{{i\pi}^{d/2}} k^{\mu} \, k^{\nu} \, \prod_{r=1}^{n} \, {c_r^{-1}}
  =  \sum_{i,j=1}^{n} \, q_i^{\mu}\, q_j^{\nu} \, n_{ij} \,  \, I_{n,ij}^{d+4} -\frac{1}{2}
   \, g^{\mu \nu}  \, I_{n}^{d+2} ,
 \\
\label{tensor3}
I_{n}^{\mu\, \nu\, \lambda}& =& \int \frac{d^d k}{{i\pi}^{d/2}} k^{\mu} \, k^{\nu} \,  k^{\lambda} \, \prod_{r=1}^{n} \, {c_r^{-1}}
  = - \sum_{i,j,k=1}^{n} \, q_i^{\mu}\, q_j^{\nu}\, q_k^{\lambda} \,  n_{ijk} \,  \, I_{n,ijk}^{d+6}
  +\frac{1}{2} \sum_{i=1}^{n} g^{[\mu \nu} q_i^{\lambda]} I_{n,i}^{d+4} ,
\label{tensor4}
\\
I_{n}^{\mu\, \nu\, \lambda\, \rho} &=&
\int \frac{d^d k}{{i\pi}^{d/2}} k^{\mu} \, k^{\nu} \,  k^{\lambda} \, k^{\rho} \, \prod_{r=1}^{n} \, {c_r^{-1}}
\nn\\
&=&
    \sum_{i,j,k,l=1}^{n} \, q_i^{\mu}\, q_j^{\nu}\, q_k^{\lambda}
 \, q_l^{\rho}\,  n_{ijkl} \,  \, I_{n,ijkl}^{d+8}
    -\frac{1}{2} \sum_{i,j=1}^{n} g^{[\mu \nu} q_i^{\lambda} q_j^{\rho]}
\, n_{ij} I_{n,ij}^{d+6}
 +\frac{1}{4} g^{[\mu \nu} g^{\lambda \rho]} I_{n}^{d+4} .
\label{tensor5}
\end{eqnarray}
Tensors like $g^{[\mu \nu} q_i^{\lambda]}$ are completely symmetrized, and  we define
 $n_{ij}={\nu}_{ij}=1+{\delta}_{ij}, ~ n_{ijk}= {\nu}_{ij}{\nu}_{ijk}, ~ {\nu}_{ijk}=1+{\delta}_{ik}+{\delta}_{jk}$ etc., and:
\begin{eqnarray}
   \label{eq:Inij}
    I_{p, \, i\,j \,k\cdots} ^{D,stu \cdots} =
 \int \frac{d^{D} k}{{i\pi}^{D/2}}
\prod_{r=1}^{n} \, \frac{1}{c_r^{1+\delta_{ri} + \delta_{rj}+\delta_{rk}+\cdots
                 -\delta_{rs} - \delta_{rt}-\delta_{ru}-\cdots}} .
\end{eqnarray}
The $I_{n-1,ab}^{\{\mu_1,\cdots\},s}$ e.g. is obtained from $I_{n}^{\{\mu_1,\cdots\}}$  by
shrinking line $s$ and raising the powers of inverse propagators $a,b$ ($s \ne a,b$);
$I_{n,ab}^{\{\mu_1,\cdots\},a}=I_{n,b}^{\{\mu_1,\cdots\}}$ .

For one-loop tensor integrals, recurrence relations have been derived allowing to  represent the scalar tensor coefficients
$I_{n,\{a\}}^D$
 in terms of scalar integrals in generic dimension $d$ and with natural indices, usually  $\nu_j=1$ \cite{Tarasov:1996br,Fleischer:1999hq}.
In the present work we need the relation to reduce index $j$ and dimension $D$ simultaneously and another
relation to reduce dimension $D$ only:
\bea
\label{eq:RR1}
  \left( \right)_n \nu_j    \left( \mathbf{j^{+}}  I_{n}^{D}\right)
&=&
 - {j \choose 0}_5 I_n^{D-2} + \sum_{k=1}^{n} {j \choose k}_n \left( \mathbf{k^{-}} I_{n}^{D-2} \right),
\\
\label{eq:RR2}
 \left( \right)_n (D+1-\sum_{i=1}^{n}\nu_i)      I_n^{{D}}
  &=&
 {{0 \choose 0}_n} I_n^{D-2}
 - \sum_{k=1}^n {0 \choose k}_n   \left( \mathbf{k^{-}} I_{n}^{D-2}  \right).
\eea
These relations hold for arbitrary index sets $\{\nu_i\}$.
The integrals $\mathbf{k^{-}} I_{n}^{D}$ and $\mathbf{j^{+}} I_{n}^{D}$ are obtained from $ I_{n}^{D}$ by replacing $\nu_k \rightarrow (\nu_k - 1)$ and $\nu_j \rightarrow (\nu_j + 1)$, respectively.
For the definitions of Gram determinant $\left( \right)_n$ and signed minors like
${j \choose k}_n$  we refer to \cite{Melrose:1965kb,Diakonidis:2008ij,Riemann:ACAT2010talkatthisconf}.
If the Gram determinant vanishes, $\left( \right)_n = 0$,  relation (\ref{eq:RR2}) allows to express $I_{n,\{a\}}^{D}$  by simpler integrals, some of them with less external legs:
\bea\label{eq:RR1a}
I_{n,\{a\}}^{D} |_{\left( \right)_n = 0} &=&
\frac{1}{{0 \choose 0}_n}
\sum_{k=1}^n {0 \choose k}_n \left( \mathbf{k^{-}} I_{n,\{a\}}^{D} \right).
\eea
All this works fine, but for tensors of rank $R$ one gets an intermediate  scalar basis with dimensions up to  $D=d+2R$, and finally relations in generic dimension $d$ with coefficients
$(1/()_n)^R$.
If $()_n$ becomes small, and this happens during phase space integrations, numerical instabilities will appear.

From other approaches it is well-known that the appearance of powers of inverse Gram determinants for 5-point functions may be avoided completely, and for 4-point functions one has to apply special measures if needed.

In \cite{Diakonidis:2008ij}, we have demonstrated for pentagons up to $R=3$ that a clever use of properties of signed minors allows to cancel all the  inverse Gram derminants  $()_5$.
In  \cite{Diakonidis:2009fx}, we derived a recursive  algorithm for the representation of $(n,R)$ tensors by  $(n,R-1)$ tensors and $(n-1,R-1)$ tensors, although with appearance of  inverse Gram determinants  $()_n$.

Here, we describe an algorithm which combines both approaches and allows an efficient evaluation of the tensor integrals in terms of scalar functions; the latter may be evaluated by packages like FF \cite{vanOldenborgh:1990yc}, LoopTools/FF\cite{Hahn:1998yk,vanOldenborgh:1990yc}, QCDloop/FF \cite{Ellis:2007qk,vanOldenborgh:1990yc}.

The algorithm  has been worked out  until $(n,R)=(6,6)$ tensors, but it is evident how to go beyond that.

\section{A sample reduction free of $1/()_5$: the 5-point tensor $I_5^{\mu\nu\lambda}$\label{s-red}}
For details we have to refer to \cite{Riemann:ACAT2010talkatthisconf} and references therein.
As an example, we consider a rank $R=3$ pentagon $I_5^{\mu\nu\lambda}$.
The rank $R=3$ tensor was treated already in \cite{Diakonidis:2008ij}, but the calculational method applied here may be more easily extended to higher ranks.
We apply the recurrence derived in \cite{Diakonidis:2010rs}:
\begin{eqnarray}
\label{tensor5general}
I_5^{\mu\nu\lambda}  &=&I_5^{\mu\nu} Q_0^{\lambda} -  \sum_{s=1}^{5}
I_4^{\mu\nu,s } Q_s^{\lambda},
\end{eqnarray}
where the auxiliary vectors contain inverse Gram determinants:
\begin{eqnarray}
\label{Q5}
 Q_s^{\mu} &=& \sum_{i=1}^{5}  q_i^{\mu} \frac{{s\choose i}_5}
  { \left(  \right)_5}
,~~~ s=0, \ldots, 5 .
\end{eqnarray}
The vector ($R=1$) is free of the inverse Gram determinant $()_5$. Starting the general recursion
\cite{Diakonidis:2009fx}, as next
the rank $R=2$ tensor is expressed by scalar 4-point functions, free of inverse $()_5$:
\begin{equation}
I_{5}^{\mu \nu}
= \sum_{i,j=1}^{4}  q_i^{\mu} q_j^{\nu}
\frac{1}{{0\choose 0}_5} \sum_{s=1}^5  \left[{0i\choose sj}_5 I_4^{d+2,s}
+
{0s\choose 0j}_5 I_{4,i}^{d+2,s} \right]
+
g^{\mu \nu}
 \left[-\frac{1}{2} \frac{1}{{0\choose 0}_5} \sum_{s=1}^5 {s\choose 0}_5 I_4^{d+2,s}\right] .
\label{final2}
\end{equation}
After further
 involved manipulations, one may arrive at an expression where also
$I_{5}^{\mu \nu \lambda}$
 is expressed by scalar 4-point functions with higher indices and in higher dimensions:
\begin{eqnarray}
I_{5}^{\mu \nu \lambda}&=& \sum_{i,j,k=1}^{4} \,
{    q_i^{\mu} q_j^{\nu}  q_k^{\lambda} }
E_{ijk}+\sum_{k=1}^4 {   g^{[\mu \nu} q_k^{\lambda]} }
E_{00k} ,
\label{Exyz0}
\\
E_{ijk}&=&-\frac{1}{{0\choose 0}_5} \sum_{s=1}^5 \left\{ \left[{0j\choose sk}_5 { I_{4,i}^{d+4,s}}
+
(i \leftrightarrow j)\right]+{0s\choose 0k}_5 {\nu}_{ij} { I_{4,ij}^{d+4,s}}
 \right\} ,
\label{Exyz2}
\nonumber \\
\\
E_{00j}&=&\frac{1}{{0\choose 0}_5} \sum_{s=1}^5 \left[\frac{1}{2} {0s\choose 0j}_5 { I_4^{d+2,s}}
- \frac{d-1}{3} {s\choose j}_5 {I_4^{d+4,s}}
 \right] .
\end{eqnarray}
The presentation is  evidently free of $1/()_5$, and it is more compact than that given in our earlier paper \cite{Diakonidis:2008ij}.
\section{The 4-point scalars and their Gram determinants\label{s-4point}}
We have now to express efficiently the following scalar functions:
\bea
I_{4}^{d+2}, I_{4}^{d+4}, I_{4,i}^{d}, I_{4,i}^{d+2}, I_{4,i}^{d+4},
 I_{4,ij}^{d}, I_{4,ij}^{d+2}, I_{4,ij}^{d+4}, \ldots
\eea
We can again treat only an example.
The recurrence relation (\ref{eq:RR1}) applies e.g.
to the scalar function $I_{4,ijk}^{d+6}$, appearing as tensor coefficient in (\ref{tensor5}):
\bea   
\label{A533}
{\nu}_{ij}{\nu}_{ijk} I_{4,ijk}^{d+6}
&=&
-\frac{{0\choose k}_4}{\left(  \right)_4} {\nu}_{ij}I_{4,ij}^{d+4}
+
 \sum_{t=1,t \ne i,j}^{4} \frac{{t\choose k}_4}{\left(  \right)_4} {\nu}_{ij}I_{3,ij}^{d+4,t}
+
 \frac{{i\choose k}_4}{\left(  \right)_4} I_{4,j}^{d+4}
+
 \frac{{j\choose k}_4}{\left(  \right)_4} I_{4,i}^{d+4}  .
\label{I4ijkd3b}
\eea
Here, we see the appearance of the inverse sub-Gram determinant ${s\choose s}_5 \equiv ()_4$.
For every dimensional shift, another inverse power of it will appear.
In contrast to the case $n=5$, this may not  be completely prevented, but the following strategy is quite useful:
\emph{Restrict the appearance of $()_4$ to terms related to $I_4^{d+2l}$, where they may be made implicit, and hold the scalar integrals $I_3,I_2,I_1$ free of them.}

A lengthy  calculation yields:
\bea
{\nu}_{ij}{\nu}_{ijk} I_{4,ijk}^{d+6}=&&
-\frac{{0\choose i}}{{0\choose 0}}
\frac{{0\choose j}}{{0\choose 0}}\frac{{0\choose k}}{{0\choose 0}}(d-1)d(d+1)I_4^{d+6}
-\frac{{0i\choose 0j}{0\choose k}+{0i\choose 0k}{0\choose j}+{0j\choose 0k}{0\choose i}}
{{0\choose 0}^2}(d-1)I_4^{d+4} \nn \\
&&+\frac{{0\choose j}}{{0\choose 0}}\frac{{0\choose k}}{{0\choose 0}}\frac{(d-1)d}{{0\choose 0}}
\sum_{t=1}^4 {0t\choose 0i}I_{3}^{d+4,t}
-\frac{{0\choose k}}{{0\choose 0}}\frac{d-1}{{0\choose 0}}
\sum_{t=1}^4 {0t\choose 0j}I_{3,i}^{d+4,t}\nn \\
&&+\sum_{t=1}^4 \frac{{0i\choose 0k}{0t\choose 0j}+
{0j\choose 0k}{0t\choose 0i}}{{0\choose 0}^2}I_{3}^{d+2,t}
+\frac{1}{{0\choose 0}}
\sum_{t=1,t \ne i,j}^4 {0t\choose 0k} {\nu}_{ij} I_{3,ij}^{d+4,t} .
\label{fulld3}
\eea
This expression is free of inverse Gram determinants $()_4$.
Although, by repeatedly applying (\ref{eq:RR2}), the expression of $I_4^{d+6},I_4^{d+4}$ in terms of  scalar integrals in generic dimension $d$ will introduce unavoidably such terms:
\bea
\label{A401}
I_{4}^{d+2l}&=&\left[\frac{{0\choose 0}_4}{\left(  \right)_4}I_{4}^{d+2(l-1)}-
\sum_{t=1}^{4} \frac{{t\choose 0}_4} {\left(  \right)_4} I_{3}^{d+2(l-1),t}  \right]
\frac{1}{d+2 l-3}.
\eea
The same recurrence relation allows to express these scalar functions $I_{4}^{d+2l}$ in terms of simpler ones at $()_4 \equiv 0$, if one writes the recurrence for $I_{4}^{d+2(l+1)}$; see (\ref{eq:RR1a}).
One may, however,  rewrite  (\ref{eq:RR2}) at arbitrary $()_4$:
\bea
\label{eq:RR1b}
 I_4^{D}
 &=&
\frac{1}{ {{0 \choose 0}_4}}
\left[
 \left( \right)_4 (D-3)      I_4^{{D+2}}
 +
 \sum_{k=1}^4 {0 \choose k}_4   I_{3}^{D,k} \right],
\eea
and apply that relation for small  $()_4$.
The $I_4^{{D+2}}$ will be evaluated (\emph{approximately}) at $()_4=0$.
This should give a better approximation than (\ref{eq:RR1a})  for the higher-dimensional functions at small  $()_4$, and may be even further iterated.

In the next section, we will study a numerical example.

\section{Numerical example: $D_{111}$ \label{s-num}}
In order to investigate the stability of the method near a typical kinematical point of vanishing sub-Gram determinant, we consider an example given
in \cite{Denner:THHH2009}, namely
the tensor integrals related to a certain box diagram, which in  LoopTools \cite{Hahn:1998yk,Hahn:2010aa} notations is:
\\
\bea
\mathrm{D0i}(\mathrm{id},0,0,s_3,s_4,s_{12},s_{23},0,M^2,0,0). 
\eea
The Gram determinant is:
\bea
()_4 ~=~ \varDelta^{(3)}=\mathrm{Det}\left( 2 K_iK_j \right),
\eea
where $K_i$ are the internal momenta, expressible by the (incoming) external momenta $p_i$: $K_1=p_1, K_2=K_1+p_2, K_3=K_2+p_3, K_4=0$.
Then, with $p_i^2=s_i, (p_i+p_j)^2=s_{ij}$, we get:
\bea
()_4 = -2 s_{12}\left[s_{23}^2 + s_3s_4 - s_{23}(s_3+s_4-s_{12}) \right] .
 \eea
This Gram determinant vanishes if:
\bea
s_4 \to s_{crit} = s_{23}\frac{(s_{23}-s_3+s_{12})}{(s_{23}-s_3)} .
\eea
For $s_{23} = 2\times 10^4$ GeV$^2$, $s_{3} = 1\times 10^4$ GeV$^2$, $s_{12} = -4\times 10^4$ GeV$^2$, it is $ s_{crit} = -6\times 10^4$ GeV$^2$, and we look at the dependence of the tensor coefficients on
\bea\label{def-x}
x = \frac{s_4}{s_{crit}} - 1 \rightarrow 0.
\eea
In these variables:
\bea
()_4  = -  2 ~x~ s_{23} s_{12} (s_{23} - s_3 + s_{12}).
\eea
In LoopTools conventions, the tensor coefficients $D_{ijl}$ are defined as follows:
\bea\label{dmunula}
D_{\mu\nu\lambda}
&=&
\sum_{i,j,l=1}^3 K_{i\mu}K_{j\nu}K_{l\lambda} D_{ijl}
+ \sum_{i=1}^3 (g_{\mu\nu}K_{i\lambda} + g_{\nu\lambda}K_{i\mu}  +g_{\lambda\mu}K_{i\nu}) D_{00i},
\eea
and for our conventions, see  (\ref{tensor3}).
The inverse propagators are $c_j=  [(k-q_j)^2-m_j^2] = [(k+K_{j-1})^2-m_j^2]$.
Because we assume in our  formulae $q_4=0$, and in LoopTools it is $K_1=0$, one has to care about specific correspondences; it is e.g.:
\bea\label{eq-i4222d111}
 D_{111} &=& n_{222}~I_{4,222}^{d+6} ,
\eea
with $n_{222}=\nu_{22}\nu_{222}=6$, and $\nu_{ij}\nu_{ijk}I_{4,ijk}^{d+6}$ given in (\ref{fulld3}).
In the example, the tensor coefficients are finite.
For tensors with  non-vanishing $1/\eps^n$ terms, there may arise modifications due to different  normalizations.

\begin{figure}[t]
\includegraphics[width=.9\textwidth]{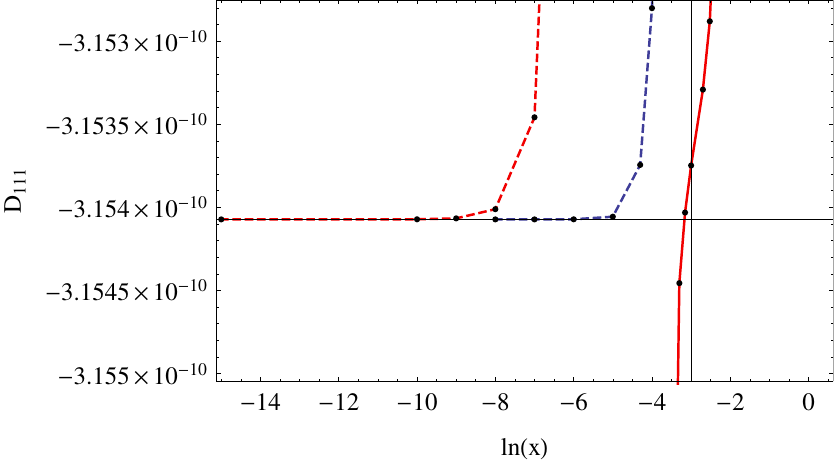}
 \caption[]{
\label{fig-L012}
The tensor coefficient $D_{111}$, defined in (\ref{dmunula}), evaluated with (\ref{eq-i4222d111}), as a function of $x$, defined in (\ref{def-x}).
Red line:  evaluation of  $D_{111}$ with (\ref{fulld3}), using reductions  (\ref{A401});
red broken line:  use of  (\ref{eq:RR1a});
blue broken line:  use of  (\ref{eq:RR1b}).
}
\end{figure}

We present  in table \ref{tab-111} and figure \ref{fig-L012}  the sample numerics for $D_{111}$, for $M=91.1876$.
Our numbers for $I_{4,222}^{d+6}$ are evaluated with a Mathematica notebook
and for comparison we also used LoopTools v.2.4,
 both approaches  in normal double precision.

\begin{table}[t]
\begin{scriptsize}
\centering
\begin{tabular}{|l|r@{.}l|r@{.}l|r@{.}l|r@{.}l|}
\hline
$x$ &  \multicolumn{2}{|c|}{ $\EuFrak{Re}$ $D_{111}^{HD}$} & \multicolumn{2}{|c|}{$\EuFrak{Im}$  $D_{111}^{HD}$} & \multicolumn{2}{|c|}{  $\EuFrak{Re}$ $D_{111}^{LT}$}
  & \multicolumn{2}{|c|}{  $\EuFrak{Im}$  $D_{111}^{LT}$}
\\
\hline
$0.0$   [0]& --{\bf 3}&{\bf 15407250453}~E-10 & --{\bf 3}&{\bf 31837792633}~E-10 &  \multicolumn{2}{|c|}{  -- }  & \multicolumn{2}{|c|}{  -- }
\\
$10^{-15}$    [0]& --3&15407250450~E-10 & --3&31837792634~E-10 &  \multicolumn{2}{|c|}{  -- }  & \multicolumn{2}{|c|}{  -- }
\\
$10^{-10}$    [0]& --3&15407189016~E-10 & --3&31837792176~E-10 &  \multicolumn{2}{|c|}{  -- }  & \multicolumn{2}{|c|}{  -- }
 \\
$ 10^{-8}$  [lin]& --3&15407250056~E-10 & --3&31837790700~E-10 &  \multicolumn{2}{|c|}{  -- }  & \multicolumn{2}{|c|}{  -- }
 \\
$ 10^{-7}$  [lin]& --3&15407246361~E-10 & --3&31837773302~E-10 &  \multicolumn{2}{|c|}{  -- }  & \multicolumn{2}{|c|}{  -- }
 \\
$ 10^{-6}$  [lin]& --3&15407198445~E-10 & --3&31837599320~E-10 &  \multicolumn{2}{|c|}{  -- }  & \multicolumn{2}{|c|}{  -- }
 \\
$ 10^{-5}$        [lin]& --3&15405621398~E-10 & --3&31835860039~E-10 &  \multicolumn{2}{|c|}{  -- }  & \multicolumn{2}{|c|}{  -- }
 \\
$5\times 10^{-5}$ [lin]& --3&15374472234~E-10 & --3&31828141644~E-10 &  \multicolumn{2}{|c|}{  -- }  & \multicolumn{2}{|c|}{  -- }
\\
$5\times 10^{-4}$  &
             --3&{\bf 1544}5603308~E-10 & --3&{\bf 3169}7733987~E-10 & --3&15444393358~E-10 & --3&31694238268~E-10
\\
$7\times 10^{-4}$  & --3&{\bf 1540}3131303~E-10 & --3&{\bf 316}83867119~E-10 & --3&15408833562~E-10 & --3&31674350685~E-10
\\
$10^{-3}$  & --{\bf 3}&{\bf 15374}819252~E-10 & --{\bf 3}&{\bf 3164}1477212~E-10 & --3&15374217567~E-10 & --3&31639655233~E-10
\\
$10^{-2}$  & --{\bf 3}&{\bf 1500800}0183~E-10 & --{\bf 3}&{\bf 2991592}0672~E-10 & --3&15007998895~E-10 & --3&29915924109~E-10
\\
$10^{-1}$  & --{\bf 3}&{\bf 1122675069}5~E-10 & --{\bf 3}&{\bf 1358233197}6~E-10 & --3&11226750694~E-10 & --3&13582331977~E-10
\\
\hline
\end{tabular}
\caption[]{\label{tab-111}$D_{111}$ calculated with the algebraic approach in comparison with LoopTools numerics.
Both calculations agree in digits shown in boldface.
Numbers labeled with [0] rely on (\ref{eq:RR1a}) with $()_4=0$, those with [lin] rely on  (\ref{eq:RR1b}), the others on  (\ref{A401}).
}.
\end{scriptsize}
\end{table}

When $x<10^{-3}$, the table shows that
both the numbers from LoopTools and from a combined use of (\ref{fulld3}) and (\ref{A401}) (in figure: red line) loose precision significantly.
With  (\ref{fulld3}) and (\ref{eq:RR1a}) we may determine the exact value of $D_{111}$ at $x=0$, and then interpolate to rising $x$ (red broken line).
The broken blue line uses  (\ref{fulld3}) and (\ref{eq:RR1b}), and it extends the region of validity of the broken red line considerably, although one does not reach the main region, where the red line is numerically safe.
A simple numerical interpolation would close the gap quite satisfactory because $D_{111}$ is a smooth function of its arguments near $x=0$.
With much more effort, namely going further with additional iterations a la (\ref{eq:RR1b}), one may reach the same.
But, one should have in mind the scales involved, and if an accuracy of, e.g., three or four digits is sufficient, simple interpolation schemes will suffice.

\section{Summary\label{s-sum}}
We gave an introduction to our purely algebraic approach to tensor reduction of one-loop Feynman integrals.
The treatment of inverse sub-Gram determinants has been refined, and a case study for vanishing $()_4$ has been presented.
For not too small Gram determinants $()_4$, the algorithm has been realized in
the Fortran package OLOTIC \cite{olotic:2010aa}, which follows the recursive approach \cite{Diakonidis:2010rs} for
tensor integrals $n\leq 6, R\leq n$.
The OLOTIC is being made an open source package.
A C++ package is under development \cite{c++yundin:2010aa} and will evaluate the tensor integrals according to the scheme described here.

\section*{Acknowledgments}
Work supported in part by Sonderforschungsbereich/Trans\-re\-gio SFB/TRR 9 of DFG
``Com\-pu\-ter\-ge\-st\"utz\-te Theoretische Teil\-chen\-phy\-sik"
and by the European Community's Marie-Curie Research Trai\-ning Network
MRTN-CT-2006-035505
``HEPTOOLS''.
We thank F. Campanario and V. Yundin for useful discussions.
J.F. likes to thank DESY for kind hospitality.

\providecommand{\href}[2]{#2}
\begingroup\raggedright
\endgroup

\end{document}